# How to Specify and How to Prove Correctness of Secure Routing Protocols for MANET
*Invited paper*

Panagiotis Papadimitratos
EPFL
Lausanne, Switzerland
panos.papadimitratos@epfl.ch

Zygmunt J. Haas
Cornell University
Ithaca, NY, USA
haas@ece.cornell.edu

Jean-Pierre Hubaux
EPFL
Lausanne, Switzerland
jean-pierre.hubaux@epfl.ch

**Abstract**

*Secure routing protocols for mobile ad hoc networks have been developed recently, yet, it has been unclear what are the properties they achieve, as a formal analysis of these protocols is mostly lacking. In this paper, we are concerned with this problem, how to specify and how to prove the correctness of a secure routing protocol. We provide a definition of what a protocol is expected to achieve independently of its functionality, as well as communication and adversary models. This way, we enable formal reasoning on the correctness of secure routing protocols. We demonstrate this by analyzing two protocols from the literature.*

## 1. Introduction

A number of protocols have been recently developed to secure the route discovery in mobile ad hoc networks, e.g, [11, 20, 18, 10]. Informally, secure routing protocols provide mechanisms that prevent adversaries, that is, nodes that deviate from the protocol definition, from influencing, controlling, or abusing the route discovery. For example, adversaries should be prevented from impersonating network destinations, advertising unreachable destinations, links that do not reflect the actual network connectivity, or misleading their peers that a destination can be reached at a lower (higher) cost than the actual one.

At first, such requirements depend on the functionality of the routing protocol. In spite of a variety of secure routing protocols for ad hoc networks proposed in the literature, there is no definition of what a protocol should achieve independently of how it operates. In other words, differing solutions have been developed without a specification.

Moreover, the requirements themselves do not capture the characteristics of the communication environment and the adversary. The literature so far assumes mostly a de facto data link layer protocol on top of which the routing protocol operates. At the same time, the capabilities of the adversaries have not been explicitly defined, either in terms of what the adversaries know or what can be their actions.

Finally, the security of different protocols has been analyzed mostly through informal arguments, with a small number of works taking a formal approach [11, 19, 3] but not addressing all the above-mentioned problems.

Our contribution here is a specification, that is, a definition of the sought properties for any candidate protocol, independently of the protocol's design and mechanisms. In particular, we define the properties of the protocol's output, one or more discovered routes. We say that a protocol is correct if it satisfies the specification. Furthermore, we define an adversary model, also independent of the protocol functionality. In addition, we outline a network communication model that captures the features of the ad hoc paradigm. With these components, we form a framework to enable formal reasoning on the properties of any secure routing protocol. To illustrate this, we analyze two secure routing protocols. Finally, we discuss related and future work.

## 2. System Model

### 2.1. Network Model

Mobile hosts move freely within some geographical area and collaboratively support the network operation, without necessarily pursuing a common objective or running the same application. The network connectivity and membership can change frequently, and so does the network area reachable by the migrating mobile hosts. Connectivity may be intermittent even when hosts are fairly stationary, e.g., when their peers alternate between 'sleep' and 'active' periods.

We define a network node as a process with (i) a unique identity $V$, (ii) a public/private key pair $E_V$, $D_V$, (iii) a module implementing the networking protocols, e.g., routing, and (iv) a module providing communication across a wireless network interface. As mobile hosts may have more than one network interface [9], more than one node may run on a mobile host. It is convenient to view mobile ad hoc networks as systems with a single node per mobile device; however, such a consideration is not necessary for the results presented here.

We focus on the network operation above the data-link layer. The node communication at the data-link layer is modeled by the following primitives and assumptions, for some radius $R$ and time $\tau$.

1. $Send_L(V, m)$: transmits message $m$ to node $V$ within radius $R$ of the transmitting node.

2. $Bcast_L(m)$: broadcasts message $m$ to all nodes within radius $R$ of the transmitting node.

3. $Receive_L(m)$: receives message $m$ transmitted by a node within radius $R$ of the receiver; $m$ is processed at a receiver $V$ if $m$ was $Bcast_L(m)$ or $Send_L(V, m)$.

4. A link $(U, V)$ exists or it is *up* when two nodes $U$ and $V$ are able to communicate directly, i.e., $U(V)$ can receive transmissions from $V(U)$. We denote any two nodes connected by an *up* link, and thus capable of bidirectional communication, as neighbors.

5. Links are either *up* or *down*, and their state does not change faster than the transmission time of a single packet.

6. The network connectivity at a particular instance in time is the graph $G$ comprising all *up* links.

7. Transmissions from $U$ are received by all nodes $V_i$ such that $(U, V_i)$ is up during the entire duration of the packet transmission.

8. Packets are delivered across an *up* link within a maximum link delay $\tau$, or they are not delivered at all. In the latter case, the delivery failure is reported to the upper layer protocol. The data-link layer handles transient network failures, it retransmits, but it does not duplicate packets.

Communication across the network is dependent on the availability of sufficient resources (bandwidth). The shared medium implies that $k$ nodes within $R$ of each other contend and obtain a portion of the bandwidth, in principle, inversely proportional to $k$. We do not assume that the network provides fairness, which is beyond the scope of this work. In general, the available bandwidth can fluctuate, be unevenly distributed among neighbors, adversaries can selfishly attempt to transmit at high rates and the network links can be congested. Such failures can either be transient and thus masked by the data link layer, or they can persist and cause messages to be dropped from the nodes' buffers, or prevent nodes from accessing the medium altogether. The latter case is equivalent to having all affected links at the *down* state.

The widely adopted IEEE 802.11 specification, without the Wireless Equivalent Protocol (WEP) security mechanism, provides bidirectional communication according to primitives 1-3 and satisfies assumptions 7 and 8 above.[1]

In this work, we are concerned with pair-wise communication across multiple wireless links between a *source* $S$ and a *destination* $T$. We denote $S$ and $T$ as the *end nodes*, and nodes that assist the $S$, $T$ communication as *intermediate nodes*. Each node in the network is equipped with a certificate; the possession of a certificate[2] does not convey authorization or does not imply trustworthiness. Rather, it is a minimum requirement for nodes to engage in secure communication and provides the means to authenticate the origin (or the relay) of network traffic.

Digital signatures provide a straightforward mechanism to authenticate each nodes to all other network nodes. Nonetheless, which cryptographic primitives (e.g., symmetric or public key) are in use, and which nodes' public keys a network node knows (or which nodes it shares symmetric keys with) are orthogonal to our discussion. These are issues related with the implementation of any candidate secure routing protocol.

### 2.2. Adversary Model

We make no assumptions on the motivation of the network nodes, which either comply with the protocol rules (benign behavior), or deviate and actively disrupt the network operation (malicious behavior). In the former case we denote a node as *correct*, while in the latter as *adversary*. Adversaries have finite processing power and cannot mount a cryptanalytic attack to compromise a private or a secret symmetric key, or invert one-way or hash functions; as a result, cryptographic primitives are assumed secure. Given the above assumptions, we consider two models of active adversaries: *independent* and *arbitrary* adversaries.

**Definition 1**: *Independent adversaries are network nodes that ignore and do not reproduce any received message that does not comply with the operation of the networking protocols, but can generate, modify or replay any other message.*

---

[1] Note that our abstraction of $R$ does not imply an idealized communication model; $R$ is a nominal range of direct wireless communication, yet this varies and depends both on the Physical Layer protocol and the Signal to Interference and Noise Ratio at the receiver.

[2] For a survey of different approaches to equip nodes with certificates in the context of ad hoc networks see [14].

Non-compliance must be explicitly defined solely with respect to the definition of the networking protocol. Any message that does not follow the expected, protocol-specific format or fails one of the employed protocol checks is deemed as non-compliant. We emphasize, however, that traffic is non-compliant if and only if the receiving node can detect that a message does not comply with the protocol; otherwise, messages that appear to be compliant, but actually are not, will be processed as compliant by the protocol (and thus by independent adversaries).

Def. 1 imposes a restriction on the actions of adversaries, in that it disallows their reproducing or modifying and relaying any non-compliant message they receive. Nonetheless, Def. 1 allows adversaries to process and thus replay or modify compliant messages or generate any message different than the non-compliant ones they received. Furthermore, it allows adversaries to act simultaneously, with their actions (attacks), in spite of the above discussed constraints, possibly having a compound effect.

Independent adversaries' behavior allows malicious behavior and extends, in a sense, benign failure models that consider node crashes, message loss (omission failures [8]), and message transmission timing failures, when a prescribed message is sent too early, or too late, or never [5]. Yet, independent adversarial behavior is protocol-aware (but *not* protocol-specific) and thus it is not out-right more general than those failure models.[3]

As it will become clear during the protocol analysis, the model of independent adversaries, with the imposed constraints, serves as a necessary condition to achieve stronger protocol properties than those achieved without the model's constraints on the adversarial behavior. In contrast to Def. 1, we have:

**Definition 2**: *Arbitrary adversaries are network nodes that can generate any message, and replay or modify any received message.*

Adversaries that mount relatively simple attacks fit in the model of independent adversaries. Consider an adversary $M$ within the transmission range of an access point or a peer that allows high bit-rate data download or video stream access. $M$ can disrupt the route discovery to ensure that no or few routes are established across its neighborhood and, consequently, little or no network bandwidth is consumed by other data flows. Another example is a node that simply discards packets to avoid depleting its own resources (battery power or CPU cycles), or an adversary that tampers with and injects forged routing information in an attempt to attract data flows and intercept sensitive information.

Independent adversarial behavior is such that it precludes any action as a consequence of or based on receipt of a non-compliant message. Within this definition, one can identify preclusion of actions that attempt to assist other adversaries mounting an attack, if, informally, one considered non-compliant traffic attributed to misbehavior. In contrast, arbitrary adversaries can perform actions that attempt to assist ongoing attacks mounted by other adversaries.

Arbitrary adversaries are more sophisticated and powerful than independent ones, having, for example, knowledge of the identities of other adversaries in the network, devoting resources (e.g., route discoveries) to establish direct and possibly private communication with other adversaries, and exchanging traffic and information about their local execution of the protocol. The model of arbitrary adversaries can encompass a range of scenarios, from a handful of attackers that collaborate in trying to defeat a network protocol security mechanism, to adversarial nodes deployed, for example, by an industrial adversary to degrade or take advantage of the services of another operator, or an enemy that hijacks nodes and injects compromised devices in a battlefield.

## 3. Routing Specification

Let $N$ be the set of network nodes and $E$ the set of unordered pairs of distinct nodes we denote as edges or *links*. A *route* is a sequence of nodes $V_i \in N$, and edges $e_{i,i+1} = (V_i, V_{i+1}) \in E$, for $0 \leq i \leq n-1$, i.e., $route = V_0, e_{0,1}, V_1, e_{1,2}, V_2, \ldots, V_{n-1}, e_{n-1,n}, V_n$. Referring to a route as a sequence of nodes implies that for any two consecutive nodes of the route $(V_i, V_{i+1}) \in E$. We call a route with $V_0 \equiv S$ and $V_n \equiv T$ an $(S,T)$-route.

The routing protocol input is a pair of nodes, $S$ and $T$. Let $t_1$ and $t_2 > t_1$ be two points in time defining a time interval $(t_1, t_2)$, with time $t_2$ the instance at which the routing protocol returns its output to $S$. When the protocol returns its output, we say that the protocol discovers a route. Depending on the output form, we distinguish two classes of route discovery: *explicit* and *implicit*.

An explicit route discovery returns a fully and clearly expressed, readily observable $(S,T)$-route, that is a $V_0, V_1, \ldots, V_{n-1}, V_n$ sequence of nodes. An implicit route discovery is a distributed computation that returns a tuple $(V_i, V_{i+1}, V_n), i = 0, \ldots, n-1$, of the form (*current node*, *relay node*, *destination*), with each $(V_i, V_{i+1}) \in E$. The $(S,T)$-route is not readily apparent, as the protocol output to $S$ is a $(V_0, V_1, V_n)$-tuple. Yet, the route is implied through a sequence of nodes $V_j$, $1 \leq j \leq n-1$, each of them also storing a $(V_j, V_{j+1}, V_n)$-tuple.

We term a protocol performing an implicit or explicit route discovery defined above as a *basic* routing protocol. A basic routing protocol provides the structure of the route but does not provide attributes of the route or its constituent nodes and edges.

Let $f : E \rightarrow M \subseteq \Re$ be a function that assigns labels,

---
[3]For example, an omission-failing node could relay a non-compliant message.

that is, real values to edges $e_{i,i+1}$. Each label $f(e_{i,i+1}) = m_{i,i+1} \in M$, which we denote as a *link metric*, provides a quantitative description of the $e_{i,i+1}$ attribute(s). For example, a metric can capture the link's reliability or resistance to failure, calculated as the fraction of the numbers of delivered over transmitted packets across the link.

The attributes of a route can be 'summarized' by the aggregate of the labels of $e_{i,i+1} \in (S,T)$-route. The aggregate value is calculated by a function $g : M \rightarrow \Re$, the *route metric* $g(m_{0,1}, \ldots, m_{n-1,n})$, whose form is protocol-dependent. The route metric can be, for example, the sum, the product, the minimum, or the maximum of the route's constituent link metrics. Moreover, we define $l_{i,i+1}$ to be the *actual* metric value for $e_{i,i+1}$, and the aggregate $g(l_{0,1}, \ldots, l_{n-1,n})$ of the actual link metrics as the *actual* route metric.

We consider an *augmented* routing protocol. The input is $S$ and $T$, and the output is an $(S,T)$-route and: (i) for explicit discovery, a sequence of labels $\{m_{0,1}, \ldots, m_{n-1,n}\}$, with one label for each link of the $(S,T)$-route, and (ii) for implicit discovery, a route metric $g(m_{0,1}, \ldots, m_{n-1,n})$ over the link metrics distributed to the $V_i \in (S,T)$-route.

We are interested in routing protocols that ensure the three properties in Def. 3 below for the discovered route(s): *loop-freedom*, *freshness*, and *accuracy*. Loop-freedom and freshness are relevant to both basic and augmented protocols, while accuracy is relevant only to augmented protocols. We term a route discovered by a basic (augmented) protocol as correct if it satisfies loop-freedom and freshness (loop-freedom, freshness, and accuracy).

**Definition 3**:

- *Loop-freedom*: an $(S,T)$-route is loop-free if it has no repetitions of nodes.

- *Freshness*: an $(S,T)$-route is fresh with respect to an interval $(t_1, t_2)$ if each of the route's constituent links is *up* at some point in time during the interval $(t_1, t_2)$.

- *Accuracy*: an $(S,T)$-route is accurate with respect to a route metric $g$ and a constant $\Delta_{good} \geq 0$ if $|g(m_{0,1}, \ldots, m_{n-1,n}) - g(l_{0,1}, \ldots, l_{n-1,n})| < \Delta_{good}$.

Loop-freedom is self-explanatory; in our context, the property implies that adversaries cannot manipulate or abuse the route discovery to create loops.

Route freshness ensures that each of the constituent links of the discovered route was *up* recently, that is, within a $(t_1, t_2)$ interval prior to the route discovery. We clarify that freshness does not guarantee that links were *up* concurrently or throughout $(t_1, t_2)$; a link could go *down* immediately after its discovery, or links could be alternately *up* and *down*, so that a route may never be intact throughout the $(t_1, t_2)$ interval. Freshness prevents the discovery of routes comprising links that existed at no point (were *down*) during the $(t_1, t_2)$ interval, and yet are 'advertised' by an adversary.

Route accuracy provides the additional assurance that the quantitative description of a route reflects its actual attributes: the route metric calculated by the protocol is within $\Delta_{good}$ from the actual route metric. $\Delta_{good}$ is a constant such that, despite malicious or benign faults that lead to inaccurate metric values, the route metric is still 'reasonably' close to the actual value and meaningful for the protocol. The protocol- and metric-specific $\Delta_{good} \geq 0$ is allowed because, even in a benign network, impairments can affect measurements and calculations for metric values. Accuracy prevents adversaries from manipulating the metric values, contributing arbitrary metric values, altering metrics provided by other intermediate nodes, and misleading end nodes into believing that a discovered route is better than it actually is.

The number of route links, or hop count, is a special case of a route attribute, with link metric values $m_{i,i+1} = 1$ for all $i = 0, \ldots, n-1$, $g()$ the sum of the $m_{i,i+1}$, and $\Delta_{good} = 0$. The hop count is trivially given by an explicit route discovery. But for an implicit discovery it can be viewed as a route attribute.

We emphasize that routes with the properties defined above are not guaranteed to be adversary-free. A secure routing protocol cannot detect an adversary that fully abides with the routing protocol, and only later, once it belongs to a utilized route, disrupts the data communication [13]. Moreover, ensuring the correctness of the discovered routes is orthogonal to the ability to actually discover one or more correct routes. Routes may not be discovered at all times due to the compound effect of adversaries' actions and network impairments. We also note that if either $S$ or $T$ is faulty, the protocol does not ensure any of the correctness properties.

## 4. Secure Routing Correctness

We analyze the Secure Routing Protocol ($SRP$), an explicit, basic protocol [11], and the augmented version of $SRP$ [16]. Using the same framework, we have also analyzed the Distance-Vector Secure Routing Protocol (DV-SRP), an implicit augmented protocol [15], and the Secure Link State Routing Protocol (SLSP), an explicit basic protocol [12]. These protocols are diverse in terms of their design and functionality (reactive vs. proactive, distance-vector vs. source routing). However, due to space limitations, we present here only the analysis of the basic and the augmented $SRP$. The definitions of the two protocols are given in the Appendices.

We assume that traffic relayed by adversaries that act as raw data (or signal) repeaters is detected, and that each node knows its neighbors (identities and keys). Protocols that bound the propagation delay (and thus transmission range and distance) for point-to-point data link transmissions can be used [2]. These protocols, as well as protocols that use geographical location information, can prevent the establishment of 'long-haul' links across the network. Such attacks, frequently denoted as 'wormhole attacks,' are not specific to the operation of particular routing protocols but rather pertain to the neighbor discovery. Secure routing protocols either include neighbor-to-neighbor authentication as part of the route discovery (e.g., [18]), or inter-operate a secure neighbor discovery protocol (e.g., [11]). At the same time, wormhole prevention protocols necessitate authentication of transmissions between neighboring nodes. We do not dwell on the specifics of neighbor-to-neighbor authentication, as cryptographic primitives and system assumptions vary.

**Lemma 1**: *Routes discovered by $SRP$ in the presence of arbitrary adversaries are loop-free.*

**Proof**: Let $M$ an arbitrary adversary that attempts to create a loop: $M$ can include its own or any other node's identity in the $NodeList$ of the $RREQ$ more than once, it can replay a $RREQ$ in an attempt to cause other nodes to re-forward $RREQ$ and thus re-append their own address, or it can receive a $RREQ$ with a loop already formed and relay it further. In all cases, the duplicate entries in $NodeList$ will be detected by $T$. Similarly, if $M$ creates or relays a $RREP$ with a loop in the *Route* list, the duplicate entries will be detected by $S$. Note that intermediate correct nodes can also detect the loops in the $RREQ$ and $RREP$ packets as they relay them; however, it is possible that all $S, T$ traffic is relayed only by adversaries. □

**Lemma 2**: *Routes discovered by $SRP$ in the presence of independent adversaries are fresh with respect to an interval $(t_1, t_2)$, where $t_1$ is the point in time at which $S$ transmitted a $RREQ$ with identifier $Q$ seeking for $T$, and $t_2$ is the point in time at which $S$ received a $RREP$ in response to the query identified by $Q$.*

**Proof**: We outline below, for easy reference, the assumptions we rely on, from the system model and the lemma statement: (a) each node can identify the source of each $Bcast_L()$ and $Send_L()$ transmission (nodes know their neighbors), (b) traffic is deemed non-compliant if it does not follow the format of a $RREQ$ or a $RREP$ and any of checks in the protocol definition fails (Apps. A, B), (c) adversarial nodes act as independent adversaries, (d) each end node knows its peer end node (identity and key) (Appendices A, B).

Let an $RREP$ carrying a $Route = \{V_{n-1} \ldots, V_2, V_1\}$ list, and an $(S, T)$-route being the $\{S, V_1, V_2, \ldots, V_{n-1}, T\}$ sequence of nodes. Let $(V_i, V_{i+1})$ be a link that was never *up* during the $(t_1, t_2)$ interval, with $V_i$ and $V_{i+1}$ either adversaries or correct nodes. We will show that no such route will be discovered (accepted) by $SRP$. An adversary can cause the inclusion of such a link in a $RREQ/RREP$.

First, consider an adversary $V_k$, $k > i+1$, tampering with the $NodeList$ of a $RREQ$ it receives, removing and/or adding one or more node identities and relaying the tampered $RREQ'$. When $T$ responds to and returns a $RREP$, $V_{i+1}$ executes $Send_L(V_i, RREP)$, as $RREP$ appears compliant with the protocol (Steps 4.1-4.3). However, $V_i$ does not $Receive_L(RREP)$, because $(V_i, V_{i+1})$ was *down* at all times during $(t_1, t_2)$, and thus $RREP$ is never received by $S$. In the special case that $V_k$ is a neighbor of $V_i$ and executes $Send_L(V_i, RREP)$ upon receiving $RREP$, $V_i$ will reject $RREP$ as non-compliant, because the node now forwarding the $RREP$ is not $V_i$'s successor along $RREP$'s *Route* (Step 4.1, (a)-(c)), and/or it did not previously relay the query that $V_i$ had $Bcast_L(RREQ)$ (Steps 2.2.4, 4.2, assumption (a)). Furthermore, if an adversary $M$ relayed a tampered $RREQ''$ such that $M \neq V_i$, $V_i \in NodeList''$, then all $V_j$ nodes and $T$ that execute $Receive_L(RREQ'')$ will discard $RREQ''$, because the last node in $NodeList''$ is not the neighbor that relays (Step 2.3.2, assumptions (a)-(c)).

Second, consider an adversary $V_k, k \leq i$, tampering with the $RREQ$ $NodeList$. Then, $V_{i+2}$ will discard $RREQ$, either because $V_{i+1}$ is not its predecessor (Step 2.2.2, (a)-(c)), or because it detected a duplicate entry in the $NodeList$ (Step 2.2.3), as it must hold that $V_k \equiv V_{i+1}$ for the adversary to avoid having the $RREQ$ discarded.

Third, consider an adversary $V_k, k \leq i$, that tampers with entries in the $RREP$ *Route* list, removing and/or adding one or more node identities, and then relaying the tampered $RREP'$. All $V_j$ intermediate nodes with $1 \leq j < k$ relay $RREP'$, since it appears compliant with the protocol. However, $S$ will discard $RREP'$, because $f_K(S, T, Q, Route') \neq f_K(S, T, Q, Route)$ (Step 4.5). Furthermore, if an adversary $M$ relayed a tampered $RREP''$ such that $M \neq V_i$ and $\forall V_i \in Route''$, then all $V_j$ nodes (and $S$) that execute $Receive_L()$ will discard $RREP''$, because their successor along the $RREP''$ *Route* is not the neighbor that relays $RREP''$ (Step 4.1, (a)-(c)). In the special case that the adversary impersonates $T$, $V_i$ discards the $RREP$ because its successor is not $T$ (Step 4.1, assumptions (a)-(c)).

Fourth, consider an adversary $V_k, k \leq i$, which, upon receipt of a $RREQ$, generates a $RREP'''$ with a $Route'''$ that includes $(V_i, V_{i+1})$, and $Send_L(V_{k-1}, RREP''')$. All $V_j, 1 \leq j < k$, intermediate nodes relay $RREP'''$, which appears to be compliant with the protocol. However, $S$ discards $RREP'''$, because $f_K(S, T, Q, Route) \neq A'''$, with $A'''$ the authenticator the adversary appended to $RREP'''$ (Step 4.5, assumption (d)).

Finally, consider an adversary $V_k, k \leq i$, transmitting a $RREP$ generated by $T$ and identified by $Q' \neq Q$. Assuming that all $(V_j, V_{j+1})$ links of the $RREP$ $Route$ are $up$ for $1 \leq j < k$, all intermediate nodes $V_j$ deem the $RREP$ compliant and relay it towards $S$. However, $S$ maintains the value of $Q$ that identifies $RREQ$ of the current route discovery (App. A). Thus, it discards any $RREP$ generated by $T$ and identified by $Q' \neq Q$ as non-compliant, because $f_K(S, T, Q, Route) \neq A' = f_K(S, T, Q', Route)$ (Step 4.5). Moreover, the adversary cannot forge a $RREQ$ from $S$ seeking for $T$ and identified by $Q$ before time $t_1$, and thus mislead $T$ to respond with a $RREP$ identified by $Q$. As a result, the adversary cannot send such a $RREP$ to $S$ after $S$ actually transmits a $RREQ$ identified by $Q$, because $T$ responds with a route reply only to $RREQ$ whose origin is $S$ (Steps 1.1.4, 2.3.4, assumption (d)). $\square$

For the augmented version of the protocol, all nodes use the same algorithm to calculate or estimate $m_{i,i+1}$ for their incident links. For $(V_i, V_{i+1})$, we denote the metric calculated by $V_i$ as $m_{i,i+1}^i$ and the one calculated by $V_{i+1}$ as $m_{i,i+1}^{i+1}$. $SRP$ requires that $m_{i,i+1}^i = m_{i,i+1}^{i+1}$ or $|m_{i,i+1}^i - m_{i,i+1}^{i+1}| < \epsilon$ for some $\epsilon > 0$, a protocol-selectable and metric-specific threshold that determines the maximum allowable discrepancy between $m_{i,i+1}^i$ and $m_{i,i+1}^{i+1}$. Despite the assumed symmetry of the link, $\epsilon$ allows for inaccuracy due to network impairments that may affect measurements necessary for the metric calculation. If the metric in use is a fixed, 'administrative' cost agreed upon between the two neighbors, then $m_{i,i+1}^i = m_{i,i+1}^{i+1}$ must hold. Metrics such as the willingness of the node to relay data, or its remaining battery power, can be determined only independently at each node and do not fit in the above definition.

If $g(m_{0,1}^1, \ldots, m_{n-1,n}^n) = \sum_{i=0}^{n-1} m_{i,i+1}^{i+1}$, we denote the function $g$ as $g_{add}$ and the constant $\Delta_{good}$ as $\Delta_{good}^{add}$, if $g(m_{0,1}^1, \ldots, m_{n-1,n}^n) = \max_{0 \leq i \leq n-1}\{m_{i,i+1}^{i+1}\}$ the function is denoted as $g_{max}$ and the constant as $\Delta_{good}^{max}$, and if $g(m_{0,1}^1, \ldots, m_{n-1,n}^n) = \min_{0 \leq i \leq n-1}\{m_{i,i+1}^{i+1}\}$ as $g_{min}$ and $\Delta_{good}^{min}$. For $m_{i,i+1}^{i+1} > 0$, $g(m_{0,1}^1, \ldots, m_{n-1,n}^n) = \prod_{i=0}^{n-1} m_{i,i+1}^{i+1}$ can be written as $g_{add}(\bar{m}_{0,1}^1, \ldots, \bar{m}_{n-1,n}^n)$, where $\bar{m}_{i,i+1}^{i+1} = \log(m_{i,i+1}^{i+1})$, for $0 \leq i \leq n-1$.

**Lemma 3**: *Routes discovered by $SRP$ in the presence of independent adversaries are accurate, with respect to (i) $g_{add}$ and $\Delta_{good}^{add} = n^2\epsilon + n\tilde{\delta}$, (ii) $g_{max}$ and $\Delta_{good}^{max} = n\epsilon + \tilde{\delta}$, and (iii) $g_{min}$ and $\Delta_{good}^{min} = n\epsilon + \tilde{\delta}$, with $n$ the number of route links, $\epsilon > 0$ the maximum allowable difference between $m_{i,i+1}^i$ and $m_{i,i+1}^{i+1}$, and $\tilde{\delta} \geq 0$ the maximum error for a metric calculation by a correct node.*

**Proof**: Consider an adversary $V_i$ that modifies one or more of the $MetricList$ entries, relaying a $RREP$ with a tampered $MetricList'$. $S$ will discard such a $RREP$, because $f_K(S, T, Q, Route, MetricList') \neq A'$. Next, consider an adversary $V_i$ that tampers with one of the values in the $MetricList$, for $j < i-1$, and relays a $RREQ$ with the tampered metric list. The $RREQ$ appears as protocol-compliant to intermediate nodes and $T$, which generates a $RREP$. When $RREP$ arrives back at $V_j$, $m_{S,i} - m'_{S,i} \neq 0$, and $V_j$ rejects $RREP$ as non-compliant.

Moreover, let $V_i$ tamper with one of the in the $MetricList$, for $j \geq i$, and relay $RREQ$ with a tampered metric list. Then, $V_{i+1}$ will reject $RREQ$ as non-compliant, because all $m_{j,j+1}^{j+1} \in MetricList$ for $j \geq i$ must be void, as these entries correspond to links not yet discovered. If $V_i$ appended one or more additional entries to $NodeList$, $RREQ$ would be discarded as $V_i$'s neighbors $Receive_L(RREQ)$ with the last node in $NodeList$ different from its precursor (Step 2.2.2).

Next, consider an adversary $V_i$ that appends an erroneous link metric, with $\delta_i \geq 0$ denoting the metric calculation error with respect to the actual link metric; $m_{i-1,i}^i = l_{i-1,i} \pm \delta_i$ and $m_{i,i+1}^i = l_{i,i+1} \pm \delta_i$. $V_{i-1}$ deems a $RREQ/RREP$ as compliant only if $V_i$ appends $m_{i-1,i}^i$ such that $|m_{i-1,i}^i - m_{i-1,i}^{i-1}| < \epsilon$. For the discovery of an $(S,T)$-route with $k$ nodes, the above inequality must be true for all $1 \leq i \leq n$. We consider the worst case, with $S$ and $T$ correct, i.e., $\delta_0 < \tilde{\delta}$ and $\delta_n < \tilde{\delta}$, and all intermediate nodes adversaries.

Then, for the first link $|m_{0,1}^0 - m_{0,1}^1| = |l_{0,1} \pm \delta_0 - (l_{0,1} \pm \delta_1)| < \epsilon \Rightarrow \delta_1 < \epsilon + \tilde{\delta}$; for the second link, $|m_{1,2}^1 - m_{1,2}^2| = |l_{1,2} \pm \delta_1 - (l_{1,2} \pm \delta_2)| < \epsilon \Rightarrow \delta_2 < \epsilon + \delta_1 < 2\epsilon + \tilde{\delta}$ and, in general, $\delta_i < i\epsilon + \tilde{\delta}$.

Similarly, for the route links in the reverse order, $|m_{n-1,n}^{n-1} - m_{n-1,n}^n| < \epsilon \Rightarrow \delta_{n-1} < \epsilon + \tilde{\delta}$, and in general $\delta_i < (n-i)\epsilon + \tilde{\delta}$. Thus, $\delta_i < \min\{i\epsilon + \tilde{\delta}, (n-i)\epsilon + \tilde{\delta}\}$ for $1 \leq i \leq n-1$. Since $\tilde{\delta}$ does not depend on $n$, $i$, and $\epsilon$, $\delta_i < \min\{i\epsilon, (n-i)\epsilon\} + \tilde{\delta}$.

Then, for $g_{add} = \sum_{i=0}^{n-1} m_{i,i+1}^{i+1} = g(l_{0,1}, \ldots, l_{n-1,n}) \pm \sum_{i=1}^{n-1} \delta_i \pm \tilde{\delta}$. The sum is bounded since each $\delta_i$ term is bounded: $\sum_{i=1}^{n-1} \delta_i < \sum_{i=1}^{n-1}(\min\{i\epsilon, (n-i)\epsilon\} + \tilde{\delta}) = \begin{cases} \epsilon \frac{n^2-1}{4} + (n-1)\tilde{\delta} & \text{if } n \text{ is odd,} \\ \epsilon \frac{n^2}{4} + (n-1)\tilde{\delta} & \text{if } n \text{ is even} \end{cases}$, and we select $\Delta_{good}^{add} = n^2\epsilon + n\tilde{\delta}$.

Then, for $g_{max}$ we get similarly: $g_{max} = \max_{0 \leq i \leq n-1}\{l_{i,i+i} \pm \delta_{i+1}\} = \begin{cases} \max_{0 \leq i \leq n-1}\{l_{i,i+1}\} + \max_{0 \leq i \leq n-1}\{\delta_i\} \\ \max_{0 \leq i \leq n-1}\{l_{i,i+1}\} - \min_{0 \leq i \leq n-1}\{\delta_i\} \end{cases}$, and select $\Delta_{good}^{max} = n\epsilon + \tilde{\delta}$. And for and $g_{min}$: $g_{min} = \min_{0 \leq i \leq n-1}\{l_{i,i+i} \pm \delta_{i+1}\} = \begin{cases} \min_{0 \leq i \leq n-1}\{l_{i,i+1}\} + \min_{0 \leq i \leq n-1}\{\delta_i\} \\ \min_{0 \leq i \leq n-1}\{l_{i,i+1}\} - \max_{0 \leq i \leq n-1}\{\delta_i\} \end{cases}$, and se-

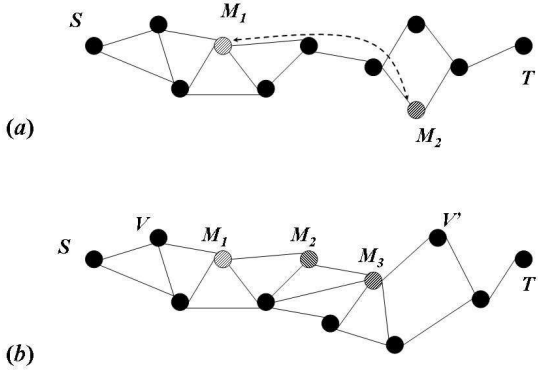

**Figure 1. Arbitrary adversaries: an illustration of two attack configurations, (a) two adversaries, (b) k=3 adversaries.**

lect $\Delta_{good}^{min} = n\epsilon + \tilde{\delta}$, to complete the proof. □

**Theorem 1**: *Routes discovered by $SRP$ in the presence of independent adversaries are loop-free, fresh, and accurate.*

**Proof**: From Lemmas 1-3. □

The assumption of independent adversaries in Theorem 1 is a necessary condition to achieve freshness and accuracy, which cannot be achieved if independence of adversaries is weakened. In the presence of arbitrary adversaries, $SRP$ provides weaker properties, discovering loop-free and *weakly fresh routes*. Informally, a route is weakly fresh if there exists a sequence of links, in general different than those comprised by the route, with each of them *up* at some point within the $(t_1, t_2)$ interval. More precisely, we call an $(S,T)$-route= $\{V_0, \ldots, V_n\}$ weakly fresh if for some $j \geq 1$, $k \leq n - 1$, and $j < k$, there exists a sequence $\{V'_0, V'_1, \ldots, V'_x\}$ of nodes such that $V'_0 \equiv V_j$, $V'_x \equiv V_k$, and all $(V'_i, V'_{i+1})$ were *up* at some point during $(t_1, t_2)$ interval, and for $i < j$ and $k \leq i < n$ all $(V_i, V_{i+1})$ were *up* at some point during $(t_1, t_2)$.

**Theorem 2**: *Routes discovered by $SRP$ in the presence of arbitrary adversaries are loop-free and weakly fresh.*

**Proof**: Lemma 1 shows loop-freedom in the presence of arbitrary adversaries. To show weak freshness, it suffices to show that at least a $(V_i, V_{i+1})$ link of the discovered route was never *up* during the $(t_1, t_2)$ due to the adversaries' actions, and then show that a sequence of links $(V'_i, V'_{i+1})$, $0 \leq i \leq x - 1$, for some $x \geq 1$ was *up* at some point during the $(t_1, t_2)$. We show two types of attacks by arbitrary adversaries leading to such a route, illustrated in Figure 1.

First, in Fig. 1.(a), let $\{S, X, M_1, Y, M_2, Z, T\}$ be an $(S,T)$-route in the network, with $X, Y \neq \emptyset, Z$, in general sequences of nodes, and $M_1, M_2$ two arbitrary adversaries. $M_1$ can implement the following protocol when receiving a $RREQ$ with $NodeList = \{X\}$:

$Send(RREQ, M_2)$; $Bcast_L(RREQ)$; Wait for $RREP$;

$M_1$ sends $RREQ$ with $NodeList = \{X, M_1\}$ directly to $M_2$, as if an $(M_1, M_2)$ link were *up*, using a $Send()$ that forwards a message across multiple hops (rather than $Send_L()$). $M_1$ can do so in a way that relaying nodes in $Y$ cannot identify the payload (e.g., by encrypting the packet). $M_1$ also broadcasts $RREQ$, so that the last node in $X$ adds $M_1$ in its $ForwardList$ and later relays the corresponding $RREP$. $M_2$ relays $RREQ$ with $NodeList = \{X, M_1, M_2\}$ and later returns $RREP$ with $Route = \{Z, M_2, M_1, X\}$ directly to $M_1$, e.g., routing the $RREP$ across $Y$.

Each of the links in $Y$ was *up* at some point in $(t_1, t_2)$, because otherwise $M_1$ would not have received the $RREP$. If $M_1$, $M_2$ were independent, $M_2$ would have ignored the $RREQ$ sent by $M_1$, as it would not be compliant. Similarly, in a variation of this attack, if $M_2$ first modified one or more entries in the $RREQ$ $NodeList$ and then $Send(RREP, M_1)$, $M_1$ would have ignored the $RREP$ for the same reason.

Second, in Fig. 1.(b), consider an $(S,T)$-route $\{X, V, M_1, M_2, \ldots, M_k, V', Y\}$, with $M_k$, $k \geq 2$, arbitrary adversaries, and $V$, $V'$ benign nodes. As long as $M_1$ and $M_k$ follow the protocol, neither $V$ nor $V'$ can discard $RREQ$ or $RREP$. However, any $M_i$, for $i \neq 1, k$, can modify $NodeList$ in an arbitrary manner and it suffices that $M_j$, for $j > i$, do not perform the checks required by the protocol and simply relay the protocol packets. In contrast, if $M_i$ were independent, $M_3$ for example would have ignored any non-compliant $RREQ$ it received from $M_2$.

In all cases, since there is at least one link that was never *up*, and adversaries can 'insert' multiple such links and contribute any arbitrary values for their link metrics, accuracy cannot be achieved. □

## 5  Related Work

The early work of [17] defined the objective of Byzantine robustness as the ability to discover a path of correct nodes, if such a path exists in the network, and proposed a reliable flooding mechanism for the dissemination of link state updates for route discovery. However, it did not provide a specification for the route discovery and the properties of the discovered routes. More recently, formal verification of distance vector protocols was considered, however, in a be-nign environment; model checking [9] and interactive theorem proving [7] techniques were used to show loop-freedom for AODV [1].

A small number of works considered formal methods and secure routing protocols for ad hoc networks. [11] analyzed $SRP$ using BAN logic, which, invented for modeling authentication protocols, lacks the expressiveness to model the operation of a routing protocol. [19] extended the Strand model [6] to allow the description of ad hoc routing protocols, and defined as goals for secure routing the ability to discover a route and then the ability to communicate across a route termed as stable. This, however, is orthogonal to the specification of the route discovery, which, as stated in [19], is not addressed.

A more recent work [3], transcribes a simulation technique previously used to prove the security of cryptographic protocols: real-world and ideal-world system models, along with two models for the adversary, one for each system, need to be defined. Then, a routing protocol is secure if the outputs of the ideal and the real-world systems are indistinguishable. In the ideal world, where essentially the adversary is thwarted, routes termed inexistent are never returned to correct nodes [3]. Nevertheless, this is not a proper definition of route properties. To illustrate this, let us assume that an existent route's links are always *up*; then, a route can exist but have loops. A more important shortcoming is revealed by the discussion of our freshness property: a route may cease to exist right after or during its discovery, or it may have never existed and yet be returned to a correct node; e.g., consider a reactive protocol wherein a link close to the destination breaks as the route reply approaches the source, or, links that are *up* only when the query/reply packets traverse them. [3, 19] assume that the topology is stable throughout the analysis.

Previous notions of adversarial models alluded to the independence definition (e.g., non-colluding nodes in [11]), or considered actions specific to a particular routing protocol functionality (e.g., [18]), or defined adversaries with respect to their physical presence and credentials they possess [10]. Each of those approaches has its merits, but our model is general enough to encompass and extend over those. Our adversary model is based on how messages are handled, and it is independent of the actual networking protocol(s) and the physical location of the code implementing them. Moreover, our reasoning does not consider only a single attacker (e.g., [3]), but allows and utilizes multiple adversaries, either independent or arbitrary.

## 6. Conclusions

The contribution of our work is to provide a framework, comprising a network model, an adversary model, and a routing specification, to enable reasoning on the correctness of secure routing protocols. We provide definitions of all three components, and analyze protocols with diverse functionality. The identification of a number of attacks demonstrates the effectiveness of our framework, which can be the basis for the analysis of any secure routing protocol. It can also be the basis for methods that seek to automate the verification of secure routing protocol properties. All these directions, along with elaborating on the adversary model and analyzing other protocols in the literature, are topics of our on-going and future research.

## APPENDIX

### A. Basic $SRP$

**Protocol Invocation**: A source node ($S$) initiates a route discovery for a destination node ($T$) only if no route discovery is under way for the same node $T$ at the time of invocation. Otherwise, a route discovery is performed at a later invocation and only after the conclusion of the ongoing route discovery. The route discovery is triggered when no $S, T$ routes are available at $S$, or it can be triggered by mechanisms independent of the routing protocol.

1. **Route Query Generation**: $S$ generates a route query or route request packet ($RREQ$).

    1.1. The route request includes the querying node $S$, the sought destination $T$, a query identifier $Q$ that was not previously used, an authenticator $A = f_K(S, T, Q)$ calculated as a function of the route query fields and a key $K$, and an empty $NodeList$.

    1.2. The node transmits the route request, i.e., $Bcast_L(RREQ)$, and it initializes a $ReplyWait$ timer.

2. **Route Query Processing**: Each node receiving a $RREQ$ determines if its own identity matches the sought destination. If not, it processes the request either as the querying node or as an intermediate node. Otherwise, it processes the request as the destination.

    2.1. **Route Query Processing at the Querying Node**:

        2.1.1. $S$ initializes an empty $ForwardList$ for each $RREQ$ it generates.

        2.1.2. $S$ adds to the $ForwardList$ each neighbor $V$ it overhears relaying $RREQ$ with $NodeList = \{V\}$.

    2.2. **Route Query Processing at Intermediate Nodes**:

        2.2.1. Each $V_k$ node invokes the $PreviouslySeen(RREQ)$ routine to specify if $RREQ$ has been previously processed. If yes, the $RREQ$ is discarded. Otherwise,

        2.2.2. $V_k$ extracts the last entry of the $NodeList$ and verifies this is the address of its precursor $V_{k-1}$.[4] If not, $RREQ$ is discarded. Otherwise,

---

[4]I.e., the node that previously $Bcast_L(RREQ)$ now processed; if $NodeList = \emptyset$, the precursor must be $S$.

- 2.2.3. $V_k$ checks the $NodeList$ for duplicate entries; if a loop is detected, $RREQ$ is discarded. Otherwise,
- 2.2.4. $V_k$ appends its own identity to the $RREQ$, updating $NodeList = \{NodeList, V_k\}$, and $Bcast_L(RREQ)$.
- 2.2.5. $V_k$ initializes an empty $ForwardList$ for each $RREQ$ it relays. It then adds to the $ForwardList$ each neighbor $V$ it overhears relaying $RREQ$ with $NodeList = \{NodeList, V\}$.
- 2.3. **Route Query Processing at Destination Node**:
  - 2.3.1. $T$ invokes the $PreviouslySeen(RREQ)$ routine to check if $RREQ$ has been previously processed. If so, the $RREQ$ is discarded. Otherwise,
  - 2.3.2. $T$ extracts the last entry of the $NodeList$, verifies that this is the address of its precursor, and discards $RREQ$ if there is a mismatch. Otherwise,
  - 2.3.3. $T$ checks if there is any duplicate entry in $NodeList$. If a loop is detected, it discards the $RREQ$; otherwise,
  - 2.3.4. $T$ calculates $f_K(S, T, Q)$ and compares it to $A$. If they are not equal, $RREQ$ is discarded; otherwise, $T$ generates and returns a route reply to $S$.
- 3. **Route Reply Generation**: $T$ generates a route reply ($RREP$).
  - 3.1. The $RREP$ packet comprises:
    - The querying node $S$
    - The destination $T$
    - The query identifier $Q$
    - A $Route$ list that contains the discovered route and also serves as the information necessary for $RREP$ to be forwarded across the network towards $S$. To determine $Route$, $T$ extracts the identifiers of the intermediate nodes previously accumulated in the $RREQ$ $NodeList$, namely, $V_1, V_2, \ldots, V_{n-1}$. $T$ stores them in reverse order in the $RREP$, setting $Route = V_{n-1}, \ldots, V_2, V_1$. And,
    - An authenticator $A' = f_K(S, T, Q, Route)$.
  - 3.2. The destination transmits the $RREP$ to the first entry of the $Route$ list: $Send_L(V_{n-1}, RREP)$.
- 4. **Route Reply Processing**:
  - 4.1. Each $V_k$, including $S$, verifies that its successor $V_{k+1}$[5] is indeed the node that now forwards the $RREP$. If not, it discards $RREP$. Otherwise,
  - 4.2. $V_k$ verifies that $V_{k+1} \in ForwardList$, unless the successor is $T$. If not, it discards $RREP$. Otherwise,
- 4.3. $V_k$ checks if there is any duplicate entry in $Route$; if yes, it discards $RREP$. Otherwise,
- 4.4. $V_k$ relays the reply to its predecessor, $V_{k-1}$, i.e., the next entry in the $Route$ list or $S$; $Send_L(V_{k-1}, RREP)$. Once $RREP$ reaches the source,
- 4.5. $S$ calculates and compares $f_K(S, T, Q, Route)$ to $A'$. If there is not a match, $S$ rejects the reply. Otherwise, it accepts the reply, and,
- 4.6. $S$ extracts the $Route$ entries to obtain the $\{S, V_1, \ldots, V_{n-1}, T\}$ route.
- 5. **Route Reply Timeout**: The $ReplyWait$ timer may expire in either of the following cases: (i) no replies from $T$, in response to the query identified by $Q$, were accepted by $S$, or, (ii) at least one reply from $T$, in response to the query identified by $Q$, was accepted by $S$. In the former case, the route discovery is considered failed, while, in the latter case, the route discovery concludes, and $S$ ignores route replies that are further delayed.
  - 5.1. **Route Discovery Failure**: $S$ initiates a new route discovery as in Step 1, using an updated value for the $ReplyWait$ timer (Step 1.2). To calculate this value between $ReplyWait_{min}$ and $ReplyWait_{max}$, $S$ invokes an $Update(ReplyWait)$ routine that returns an equal or higher value than the one previously used for the failed route discovery.
  - 5.2. **Route Discovery Conclusion**: Upon accepting a $RREP$ from $T$ identified by $Q$, $S$ considers the discovery concluded after at least $ReplyWait_{min}$ seconds elapse from the corresponding query generation, allowing then for a new route discovery, if necessary. If so, the $ReplyWait$ timer is reset, and $S$ invokes $Update(ReplyWait)$ to select $ReplyWait_{min}$ as the new route discovery timer value (Step 1.2).

**Definition 3**: *A route discovery is the current route discovery during the period of time that elapses from the generation of the route query (Step 1) till the earlier of the following two events: the expiration of the ReplyWait timer (Step 5.1 and 5.2), or a route rediscovery (Step 5.2).*

## B. Augmented $SRP$

The following steps are those that are different or added to the functionality of the basic $SRP$ defined above.

1. **Route Query Generation**:
   - 1.1. The route request includes the querying node $S$,..., an an empty $NodeList$, and an empty $MetricList$.
2. **Route Query Processing**:
   - 2.1. **Route Query Processing at the Querying Node**:
     - 2.1.1. Each neighbor $V$ that processes and is overheard forwarding a $RREQ$ with $NodeList = \{V\}$ and $MetricList = \{m_{0,1}^1\}$ is added to the $ForwardList$ if and only if $|m_{0,1}^0 - m_{0,1}^1| < \epsilon$.

---
[5] I.e., the node entry prior to $V_k$ in the $RREP$ $Route$ list, or $T$ if $V_k$ is the first entry in $Route$.

2.2. **Route Query Processing at Intermediate Nodes**:

  2.2.4. .a (before 2.2.4.) $V_k$ checks if the number of entries in the $MetricList$ is equal to the number of entries in the $NodeList$. If not, it discards the $RREQ$. Otherwise,

  2.2.4. $V_k$ appends its own identity to the $RREQ$, $NodeList$, it appends $m_{k-1,k}^k$ to the $MetricList$ and $Bcast_L(RREQ)$.

  2.2.5. $V_k$ initializes an empty $ForwardList$ for each $RREQ$ it relays. Each neighbor $V_{k+1}$ that is overheard relaying $RREQ$ with $NodeList = \{NodeList, V_{k+1}\}$ and $MetricList = \{MetricList, m_{k,k+1}^{k+1}\}$ is added to $ForwardList$ along with $m_{k,k+1}^{k+1}$ if and only if $|m_{k,k+1}^k - m_{k,k+1}^{k+1}| < \epsilon$.

  2.2.7. $V_k$ stores $m_{S,k}$, the route prefix metric calculated from the $RREQ$ $MetricList$.

2.3. **Route Query Processing at Destination Node**:

  2.3.4. .a (before 2.3.4) $T$ checks if the number of entries in the $MetricList$ is equal to the number of entries in the $NodeList$; if not, it discards the $RREQ$.

3. **Route Reply Generation**: the destination $T$ generates a route reply ($RREP$) packet comprising:

   - ...
   - $MetricList$ containing all the link metrics accumulated in the $RREQ$ along with $m_{n-1,n}^n$ (note: $m_{n-1,n}^n \equiv m_{n-1,T}^T$). Link metrics are also reversed, in order to correspond to the $RREP$ $Route$ entries.
   - $A' = f_K(S, T, Q, Route, MetricList)$.

4. **Route Reply Processing**:

  4.2. $V_k$ verifies that $V_{k+1} \in ForwardList$, unless the successor is $T$.

    4.2.1. If $V_k$ is $T$'s predecessor (i.e., $k = n - 1$), it checks whether $|m_{k,T}^T - m_{k,T}^k| < \epsilon$. If not, it discards $RREP$. Otherwise,

    4.2.2. $V_k$ checks if $m_{S,k} = m'_{S,k}$, where $m'_{S,k}$ is the aggregate value calculated from the link metric values reported in the $RREP$ $Route$ for links $(V_k, V_{k+1})$, $k < i$. If not, it discards $RREP$.

  4.5. $S$ calculates and compares $f_K(S, T, Q, Route, MetricList)$ to $A'$. If there is a match, $S$ accepts the reply, and rejects it otherwise.